\documentclass[conference]{IEEEtran}
\IEEEoverridecommandlockouts
\usepackage{cite}
\usepackage{amsmath,amssymb,amsfonts}
\usepackage{algorithmic}
\usepackage{graphicx}
\usepackage{textcomp}
\usepackage{xcolor}
\usepackage{multirow}
\def\BibTeX{{\rm B\kern-.05em{\sc i\kern-.025em b}\kern-.08em
    T\kern-.1667em\lower.7ex\hbox{E}\kern-.125emX}}

\begin{document}

\title{Compound Prompt Constraints in LLM Code Generation: A Factorial Study of Format, Persona, and Urgency}

\author{\IEEEauthorblockN{Shrenik Jadhav}
\IEEEauthorblockA{\textit{Doctoral Student} \\
\textit{Department of EECS}\\
\textit{Embry-Riddle Aeronautical University}\\
Daytona Beach, FL, USA \\
jadhavs2@my.erau.edu}
\and
\IEEEauthorblockN{Nickalas La Placa}
\IEEEauthorblockA{\textit{Undergraduate Research Assistant} \\
\textit{Department of EECS}\\
\textit{Embry-Riddle Aeronautical University}\\
Daytona Beach, FL, USA \\
laplacan@my.erau.edu}
\and
\IEEEauthorblockN{Caleb Stone}
\IEEEauthorblockA{\textit{Graduate Research Assistant} \\
\textit{Department of EECS}\\
\textit{Embry-Riddle Aeronautical University}\\
Daytona Beach, FL, USA \\
stonec16@my.erau.edu}
\and
\IEEEauthorblockN{Ashok Raja}
\IEEEauthorblockA{\textit{Assistant Professor} \\
\textit{Computer Information Technology and Graphics}\\
\textit{Purdue University Northwest}\\
Hammond, IN, USA \\
raja22@pnw.edu}
\and
\IEEEauthorblockN{Omar Ochoa}
\IEEEauthorblockA{\textit{Associate Professor} \\
\textit{Department of EECS}\\
\textit{Embry-Riddle Aeronautical University}\\
Daytona Beach, FL, USA \\
ochoao@erau.edu}
\and
\IEEEauthorblockN{Vidhyashree Nagaraju}
\IEEEauthorblockA{\textit{Assistant Professor} \\
\textit{Department of EECS}\\
\textit{Embry-Riddle Aeronautical University}\\
Daytona Beach, FL, USA \\
Vidhyashree.Nagaraju@erau.edu}
}

\maketitle

\begin{abstract}
Large language models (LLMs) are increasingly deployed in software engineering pipelines for code generation, where production prompts routinely combine multiple constraints in a single request. A typical prompt may require the response to follow a structured output format such as JSON or XML, may assign the model a specific developer persona through the system message, and may frame the request with urgency or quality language reflecting real-world deployment pressures. Prior work has studied each of these prompt-level interventions in isolation, but the interaction between them remains uncharacterized, leaving open whether single-factor evaluations can predict the behavior of the compound prompts that real systems actually issue. To address this gap, this paper presents a full-factorial empirical study of how output formatting, persona assignment, and urgency framing jointly affect the reliability of LLM code generation. Rather than evaluating each constraint separately, the framework manipulates all three factors simultaneously in a controlled $3 \times 3 \times 3$ design and decomposes each compound condition into an additive prediction and a residual interaction term that quantifies super-additive degradation.

The framework is evaluated on the 164 problems of HumanEval+ across five OpenAI models spanning the GPT-4o family, the GPT-4.1 family, and the o3-mini reasoning model, yielding 22,140 individual evaluations under greedy decoding. A format-aware code extraction pipeline with multi-strategy fallback separates formatting failures from reasoning failures, and statistical significance is assessed using McNemar's test on per-problem outcomes with odds ratios and 95\% confidence intervals. The results show that compound prompt constraints produce architecture-dependent degradation that cannot be predicted from single-factor experiments. The GPT-4o family exhibits a consistent super-additive pattern in which compound constraints reduce pass@1 by 3 to 12 percentage points beyond what the sum of individual effects would predict, with the largest interaction reaching $-12.2$ pp on GPT-4o-mini for a JSON plus expert persona plus moderate urgency combination, and JSON-based combinations producing substantially larger interaction effects than XML-based combinations. In contrast, the GPT-4.1 family resists the effect at both parameter scales, and the o3-mini reasoning model exhibits a qualitatively different pattern in which structured output constraints improve rather than degrade performance. These findings indicate that the vulnerability is architecture-dependent rather than size-dependent, that individually neutral or beneficial constraints can combine to produce large degradation invisible to single-factor evaluation, and that compound-prompt testing should be a standard part of reliability assessment for LLM-assisted engineering pipelines.
\end{abstract}

\begin{IEEEkeywords}
large language models, prompt engineering, code generation, reliability evaluation, factorial experiment
\end{IEEEkeywords}

\section{Introduction}

Large Language Models (LLMs) have rapidly become a core component of modern software engineering practice. Beyond software engineering, LLMs are also being integrated into other engineering domains as decision-support and control components, including fairness-aware reward shaping in multi-agent reinforcement learning for peer-to-peer energy markets~\cite{jadhav2025fairmarket, jadhav2026scalablefairness} and explainable reinforcement learning for voltage control in power distribution networks~\cite{jadhav2026xrlllm}. As LLMs take on increasingly consequential roles across these domains, understanding how prompt construction affects output reliability becomes a cross-cutting concern. Tools built on models such as the GPT-4 family now assist developers with code completion, function synthesis, test generation, refactoring, and bug localization, and recent surveys document hundreds of studies applying LLMs across nearly every phase of the software lifecycle~\cite{jiang2024codellmsurvey, zhang2023llm4se}. Their appeal lies in a deceptively simple interface: a developer writes a natural language prompt and receives executable code in return. Behind this interface, however, the mapping from prompt to output is highly sensitive to how the request is phrased, and small changes in wording, structure, or framing can produce measurable shifts in correctness.

A growing body of work shows that prompt design is not a cosmetic concern but a first-order determinant of LLM behavior. Three categories of prompt-level interventions have received particular attention. First, structured output constraints such as JSON or XML schemas are routinely imposed so that downstream pipelines can parse model responses programmatically, yet recent work reports that requiring strict output formats can measurably degrade reasoning quality on the underlying task~\cite{tam2024format}. Second, persona or role assignment via the system message is widely used to specialize model behavior, with mixed evidence on whether expert personas reliably improve task performance~\cite{salewski2023incontext, zheng2024helpful}. Third, urgency and emotional framing has been shown to shift LLM outputs in measurable ways, sometimes improving accuracy on benchmark tasks~\cite{li2023emotional}. Each of these interventions corresponds to a common deployment pattern. Production prompts in code generation pipelines routinely combine a structured output schema, a role specification, and a quality or urgency framing in a single request, yet they are almost never studied in combination.

Despite the practical relevance of compound prompts, the published literature has overwhelmingly evaluated these factors one at a time. Studies of structured output~\cite{tam2024format}, persona assignment~\cite{salewski2023incontext, zheng2024helpful}, and urgency framing~\cite{li2023emotional} each report effects in isolation, but do not characterize what happens when all three are imposed simultaneously. This leaves a concrete methodological gap. If the individual effects were additive, single-factor results would be sufficient to predict compound behavior. If they are not additive, single-factor evaluation is systematically misleading for the prompts that real systems actually issue, and any reliability claim derived from such evaluations may not transfer to deployment. The gap is not only quantitative but also architectural: it is unknown whether sensitivity to compound constraints is a property of model size, training procedure, or model family, which has direct implications for how practitioners should select models for constraint-heavy pipelines. In deployed software systems where LLM outputs feed automated test suites, code review tools, or production refactoring workflows, an unmodeled interaction between routine prompt elements becomes a latent reliability risk, since failures appear only under the combined conditions that real production prompts actually issue.

To address this gap, this paper presents a full-factorial empirical study of how three categories of prompt constraints jointly affect the reliability of LLM code generation. Rather than evaluating output formatting, persona assignment, and urgency framing in isolation, the study manipulates all three factors simultaneously in a controlled $3 \times 3 \times 3$ design and measures how their interaction shapes per-problem pass rates across multiple model families. This design enables direct quantification of compound effects that cannot be recovered from single-factor experiments, and isolates whether observed degradation is driven by individual constraints, by their interaction, or by underlying model architecture. The framework is intended to support component-level reliability assessment of LLM-assisted code generation pipelines, where production prompts routinely combine structured output schemas, role specifications, and quality or urgency framing in a single request. 
The main contributions of this paper are as follows.
\begin{itemize}
    \item We introduce a $3 \times 3 \times 3$ full-factorial experimental framework that varies output format, persona, and urgency framing across the 164 problems of HumanEval+~\cite{liu2024evalplus, chen2021codex} and five OpenAI models spanning the GPT-4o family, the GPT-4.1 family, and the o3-mini reasoning model, yielding 22{,}140 individual evaluations under greedy decoding.
    \item We propose an interaction analysis methodology that decomposes each triple-constraint condition into an additive prediction and a residual interaction term, supported by a format-aware code extraction pipeline that separates formatting failures from reasoning failures and by McNemar paired tests with odds ratios and 95\% confidence intervals on per-problem outcomes.
    \item We empirically identify a super-additive degradation pattern in the GPT-4o family, in which compound constraints reduce HumanEval+ pass@1 by 3 to 12 percentage points beyond what the sum of individual effects would predict, with the largest interaction reaching $-12.2$ pp on GPT-4o-mini and JSON-based combinations producing substantially larger interaction effects than XML-based combinations.
    \item We show that this vulnerability is architecture-dependent rather than size-dependent. The GPT-4.1 family resists the effect at both parameter scales, and the o3-mini reasoning model exhibits a qualitatively different pattern in which structured output constraints improve rather than degrade performance.
    \item We demonstrate that single-factor prompt evaluation is insufficient for predicting compound-prompt behavior, since individually neutral or beneficial constraints can combine to produce large degradation that is invisible without factorial testing, with implications for the reliability assessment of LLM-assisted engineering pipelines.
\end{itemize}

\noindent The framework is evaluated across five models on 27 prompt conditions, producing a per-condition pass@1 surface that supports main-effect analysis across factor levels, additive interaction decomposition for all eight triple-constraint combinations, and pairwise McNemar significance testing on per-problem outcomes. Beyond the headline interaction findings, the analysis quantifies how individual constraints can be neutral or even beneficial in isolation while their combination produces large degradation, characterizes the JSON versus XML asymmetry as a practical mitigation lever, and tracks prompt token length as a confound inherent to compound-constraint research. Together, these results indicate that compound-prompt evaluation should be a standard component of reliability assessment for LLM-assisted engineering workflows, and that model selection, not just prompt design, is a critical factor in deployment reliability.

The remainder of this paper is organized as follows. Section~II introduces the background on large language models, prompt constraints, and the statistical methods used in the analysis. Section~III describes the factorial design, models, benchmark, extraction pipeline, and evaluation procedures. Section~IV presents results across all five models, decomposes the three-way interaction, and discusses implications for LLM-assisted software engineering. Section~V concludes.

\section{Background}\label{sec:background}
This section briefly introduces the core techniques and concepts used in the proposed experimental framework, focusing on their role in eliciting code from large language models and quantifying the reliability of the resulting outputs.

\subsection{Large Language Models for Code Generation}
Large language models are neural networks based on the Transformer architecture~\cite{vaswani2017attention} that are trained on large corpora of natural language and source code to learn statistical regularities of both. Code-capable LLMs such as the GPT-4 family generate programs autoregressively, predicting one token at a time conditioned on a prompt that typically contains a task description, a function signature, and optional contextual instructions~\cite{chen2021codex, jiang2024codellmsurvey}. In this framework, LLMs are treated as black-box code generators whose only controllable inputs are the system message, the user message, and the decoding parameters, mirroring the interface available to practitioners deploying these models in production pipelines.

\subsection{Prompt Constraints}
A prompt constraint is any instruction in the system or user message that restricts or shapes the model's output beyond the core task description. Three categories of constraints are central to this work. Output format constraints require the response to conform to a specific structure such as JSON or XML, so that downstream systems can parse the result programmatically; recent evidence indicates that such constraints can degrade reasoning quality when enforced through prompt instructions rather than decoding-time mechanisms~\cite{tam2024format}. Persona constraints assign a role or identity to the model via the system message, such as a generic developer or a domain expert, with the goal of specializing behavior~\cite{salewski2023incontext, zheng2024helpful}. Urgency framing prepends emotional or motivational language to the user message, such as statements describing high stakes or time pressure, which has been shown to measurably alter LLM outputs~\cite{li2023emotional}. In this framework, these three categories define the three factors of the factorial design.

\subsection{HumanEval+ and the pass@$k$ Metric}
HumanEval is a benchmark of 164 hand-written Python programming problems, each consisting of a function signature, a docstring, and a hidden test suite~\cite{chen2021codex}. A model's solution is considered correct only if the generated function passes all hidden tests when executed in a sandbox. The standard metric is pass@$k$, the probability that at least one of $k$ independently sampled solutions passes all tests:
\begin{equation}\label{eq:passk}
\text{pass@}k = \mathbb{E}_{\text{problems}} \left[ 1 - \frac{\binom{n - c}{k}}{\binom{n}{k}} \right]
\end{equation}
where $n$ is the number of samples drawn per problem and $c$ is the number of correct samples. HumanEval+~\cite{liu2024evalplus} augments each original problem with approximately 764 additional test cases generated through mutation and differential testing, substantially reducing the false-positive rate of base HumanEval and providing greater discriminative power for detecting subtle correctness failures. In this work, all conditions are evaluated using pass@1 on HumanEval+ under greedy decoding, which corresponds to a single deterministic sample per problem.

\subsection{Factorial Design and Interaction Effects}
A full-factorial design varies multiple independent factors simultaneously across all combinations of their levels, enabling estimation of both main effects, the average influence of each factor, and interaction effects, the extent to which the joint influence of multiple factors deviates from the sum of their individual effects~\cite{montgomery2017doe}. For three factors with three levels each, the design contains $3 \times 3 \times 3 = 27$ conditions. The three-way interaction term is defined as the residual between the actual outcome under the full combination and the prediction obtained by adding the individual main effects to the baseline:
\begin{equation}\label{eq:interaction}
I_{f,p,u} = Y_{\text{actual}} - \left( Y_{\text{base}} + \Delta_f + \Delta_p + \Delta_u \right)
\end{equation}
where $\Delta_f$, $\Delta_p$, and $\Delta_u$ are the main effects of the format, persona, and urgency factors measured in isolation. A negative value of $I_{f,p,u}$ indicates super-additive degradation, meaning the compound constraint produces worse performance than the sum of its parts would predict. In this framework, the interaction term is the primary quantity used to detect compound-prompt failure modes.

\subsection{McNemar's Test for Paired Binary Outcomes}
Because the same 164 HumanEval+ problems appear in every experimental condition, comparing two conditions yields paired binary outcomes for each problem, namely pass or fail under each condition, rather than independent samples. McNemar's test is the appropriate significance test for this setting~\cite{mcnemar1947note}. Given the discordant pair counts $b$, the number of problems passed under condition A but failed under B, and $c$, the number passed under B but failed under A, the test statistic and the corresponding odds ratio are:
\begin{equation}\label{eq:mcnemar}
\chi^2 = \frac{(b - c)^2}{b + c}, \qquad \text{OR} = \frac{b}{c}
\end{equation}
An odds ratio greater than one indicates that problems solved under condition A are more likely to fail under condition B than the reverse. In this work, McNemar's test is applied with 95\% confidence intervals on the odds ratio to assess the statistical significance of degradation between single-constraint and compound-constraint conditions.

\section{Methodology}

\begin{figure*}[!h]
\centering
\includegraphics[width=\linewidth]{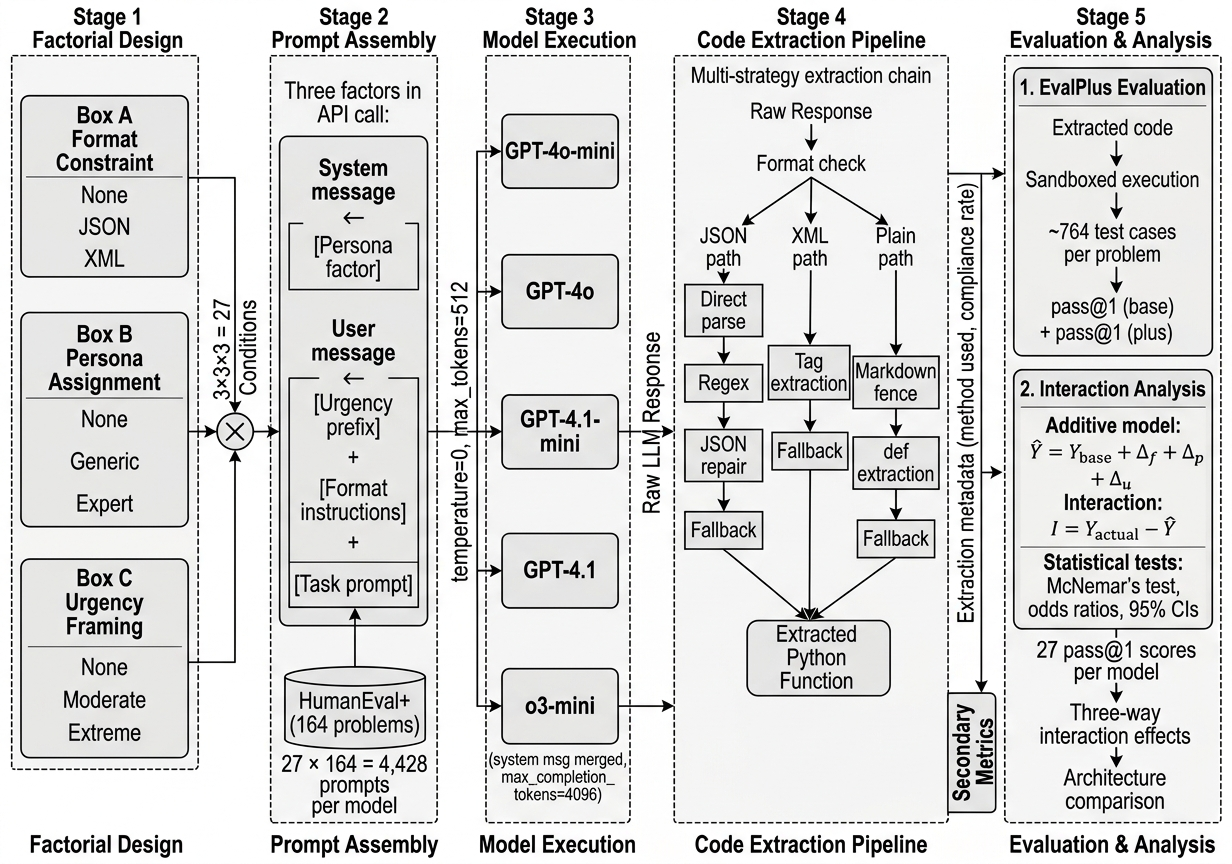}
\caption{Experimental framework. Stage~1 defines the $3 \times 3 \times 3$ factorial design. Stage~2 assembles prompts by mapping factor levels to system and user messages. Stage~3 dispatches prompts across five models. Stage~4 applies format-aware code extraction with multi-strategy fallback. Stage~5 evaluates extracted code via EvalPlus and computes three-way interaction effects.}
\label{fig:framework}
\end{figure*}

We employ a full-factorial experimental design to systematically evaluate how three categories of prompt constraints interact to affect LLM code generation reliability. Fig.~\ref{fig:framework} presents the five-stage experimental pipeline, from factorial design through evaluation and statistical analysis. The complete framework processes 22,140 individual evaluations (27 conditions $\times$ 164 problems $\times$ 5 models).

\subsection{Factorial Design}

We manipulate three independent prompt constraint factors, each at three levels of increasing intensity, producing $3 \times 3 \times 3 = 27$ experimental conditions:

\paragraph{Format Constraint} The format factor controls the structured output requirement at three levels: no formatting requirement, in which the model returns free-form code; JSON, in which the response must be a JSON object containing the solution as a string value; and XML, in which the response must be wrapped in solution and code tags. Format compliance is enforced through prompt instructions only, not through API-level mechanisms such as the response format parameter, ensuring that the model must simultaneously manage both reasoning and formatting.

\paragraph{Persona Assignment} The persona factor controls role-based identity framing delivered through the system message at three levels: no system message, a generic developer persona, and a detailed senior Python developer persona specifying 15 years of experience with algorithmic specialization and edge-case expertise.

\paragraph{Urgency Framing} The urgency factor controls emotional and motivational pressure prepended to the user message at three levels: no framing, moderate importance framing requesting thoroughness and accuracy, and extreme emergency framing describing a production outage with explicit job-loss pressure.

\subsection{Prompt Construction}

Each prompt is assembled by combining the three factor levels into a structured API call. The persona level determines the system message content, which is left empty when no persona is assigned. The user message is composed by concatenating the urgency prefix, format-specific instructions, and the target function signature from the benchmark. This design ensures that constraint complexity increases with factor intensity while maintaining consistent task semantics across all conditions.

\subsection{Models}

We evaluate five OpenAI models spanning two architectural families and one reasoning model.

The GPT-4o family includes GPT-4o-mini and GPT-4o, representing a smaller and larger variant of the same architecture. The GPT-4.1 family includes GPT-4.1-mini and GPT-4.1, a newer family explicitly optimized for improved instruction following~\cite{openai2025gpt41}. We also evaluate o3-mini, a reasoning model that generates hidden chain-of-thought tokens before producing visible output.

All standard models use greedy decoding (temperature = 0) with a maximum output length of 512 tokens. The o3-mini model requires modified parameters: system messages are merged into the user message as o3-mini does not support the system role, the maximum completion token budget is set to 4096 to accommodate hidden reasoning tokens, and temperature is fixed at 1, the only value supported by reasoning models. These differences are noted when interpreting o3-mini results.

\subsection{Benchmark} 

We use HumanEval+~\cite{liu2024evalplus}, an enhanced version of OpenAI's HumanEval benchmark~\cite{chen2021codex}. HumanEval+ contains 164 function-level Python programming problems, each augmented with approximately 764 test cases compared to roughly 10 in the original. This substantially more rigorous test suite reduces the false-positive rate inherent in base HumanEval and provides greater discriminative power for detecting constraint-induced degradation.

\subsection{Code Extraction Pipeline}

A critical challenge in evaluating structured output conditions is extracting executable code from format-wrapped responses. We implement a multi-strategy extraction pipeline that routes responses based on the expected format.

For JSON conditions, extraction attempts direct JSON parsing first, then regex-based key matching, then automated JSON repair for malformed responses, and finally falls back to plain-text extraction. A key implementation detail is handling double-escaped newline characters, which appear when models embed code within JSON strings and cause near-zero pass rates if left uncorrected.

For XML conditions, extraction searches for code content within the expected tag structure, applies HTML entity decoding, and falls back to plain-text extraction if tags are missing or malformed.

For free-form conditions, extraction checks for markdown code fences, then attempts to identify function definitions by indentation structure.

Each extraction records the method used, enabling post-hoc separation of formatting failures from reasoning failures as recommended for structured output evaluation~\cite{tam2024format}.

\subsection{Evaluation Metrics and Statistical Analysis}

Our primary metric is pass@1 on the HumanEval+ test suite, defined as the fraction of problems for which the extracted code passes all augmented test cases in a single attempt.

To quantify compound constraint effects, we compute the three-way interaction term $I$ for each triple combination as defined in Eq.~\ref{eq:interaction} of Section~\ref{sec:background}. A statistically significant negative interaction ($I < 0$) indicates super-additive degradation, in which compound constraints reduce performance beyond the sum of their individual effects.

Because the same 164 problems appear in every condition, we use McNemar's test for paired binary outcomes rather than independent-sample tests. We report odds ratios with 95\% confidence intervals as effect size measures alongside p-values from the McNemar tests.

Prompt length measured in tokens necessarily increases with constraint intensity, as format instructions and urgency framing add tokens to the prompt. We track prompt token counts across conditions and discuss this as a limitation, noting that this confound is inherent to the phenomenon under study since real-world compound prompts are longer precisely because they contain more constraints.

\section{Results and Discussion}

We evaluated five OpenAI models across all 27 prompt conditions using the HumanEval+ benchmark (164 problems, $\sim$764 test cases per problem). Each condition received a single greedy-decoded response (temperature$=$0) per problem, yielding 22,140 total evaluations. We report HumanEval+ pass@1 as the primary metric throughout, as the enhanced test suite provides substantially greater discriminative power than base HumanEval. Table~\ref{tab:summary} summarizes performance across all five models.

\begin{table}[t]
\caption{Summary of Results Across Five Models (HumanEval+ pass@1)}
\label{tab:summary}
\centering\footnotesize
\begin{tabular}{lcccccc}
\hline
\textbf{Model} & \textbf{Base.} & \textbf{Best} & \textbf{Worst} & \textbf{Range} & \textbf{Avg.Int.} & \textbf{SA} \\
\hline
4o-mini & 73.8 & 84.8 & 70.1 & 14.6 & $-$7.6 & 8/8 \\
4o & 79.3 & 87.8 & 75.0 & 12.8 & $-$4.2 & 5/8 \\
4.1-mini & 87.2 & 90.2 & 83.5 & 6.7 & $+$1.7 & 0/8 \\
4.1 & 89.6 & 90.9 & 84.8 & 6.1 & $+$1.3 & 2/8 \\
o3-mini & 59.8 & 91.5 & 59.8 & 31.7 & \textemdash & \textemdash \\
\hline
\multicolumn{7}{l}{\scriptsize Base.=Baseline (\%), Range=Best$-$Worst (pp),} \\
\multicolumn{7}{l}{\scriptsize Avg.Int.=Avg.\ three-way interaction (pp), SA=Super-additive count.} \\
\multicolumn{7}{l}{\scriptsize o3-mini excluded from interaction analysis (see Section~\ref{sec:o3}).} \\
\end{tabular}
\end{table}

\subsection{Performance Under Compound Constraints}

\begin{figure}[t]
\centering
\includegraphics[width=\columnwidth]{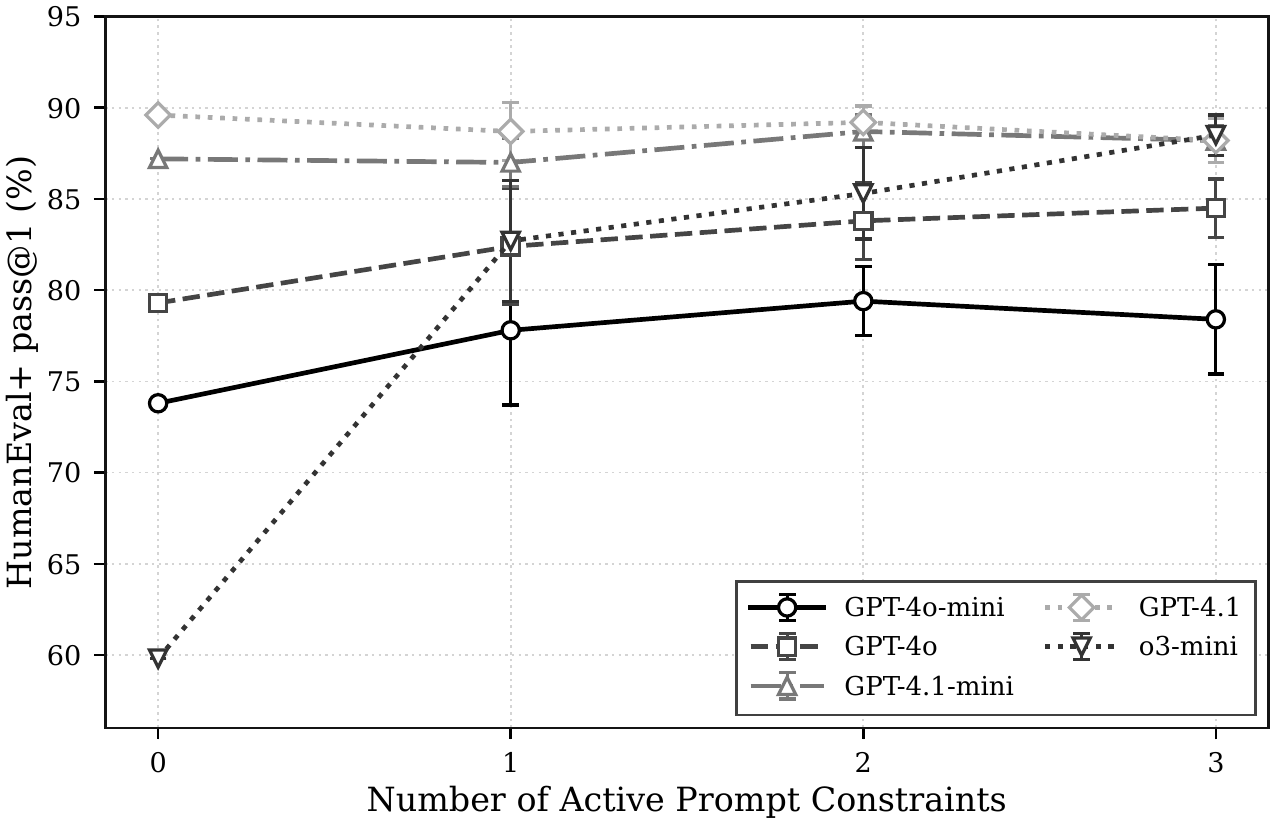}
\caption{Mean HumanEval+ pass@1 as a function of the number of active prompt constraints across five models. Error bars indicate 95\% confidence intervals. The GPT-4o family shows greater sensitivity to constraint accumulation than the GPT-4.1 family.}
\label{fig:degradation}
\end{figure}

Fig.~\ref{fig:degradation} presents the mean pass@1 as a function of the number of simultaneously active prompt constraints, ranging from baseline with no active constraint to the full triple combination. Two distinct behavioral clusters emerge. The GPT-4.1 family (4.1-mini baseline 87.2\%, 4.1 baseline 89.6\%) maintains near-flat performance across all constraint levels, with a total range of only 6.1 to 6.7 percentage points (pp) across the 27 individual conditions, as summarized in Table~\ref{tab:summary}. In contrast, the GPT-4o family exhibits substantially greater variability across the same set of conditions: GPT-4o-mini spans 14.6 pp from 70.1\% to 84.8\%, and GPT-4o spans 12.8 pp from 75.0\% to 87.8\%. The narrower ranges visible in the marginal main-effect averages of Table~\ref{tab:main_effects} reflect the smoothing effect of averaging across the other two factors and should not be interpreted as the per-condition extremes.

Notably, individual constraints do not uniformly degrade performance. Table~\ref{tab:main_effects} shows that for GPT-4o-mini, urgency framing improves average pass@1 from 76.6\% (none) to 80.2\% (extreme), and XML formatting improves it from 78.1\% (none) to 82.0\%. These individual gains are consistent with prior findings that mild emotional stimuli and structured output instructions can sharpen model attention~\cite{li2023emotional, tam2024format}. The persona factor is the clear exception: across all four standard models, persona assignment produces negligible main effects, with marginal pass@1 differences of less than 1.7 pp between the no-persona, generic-developer, and expert-persona conditions in every row of Table~\ref{tab:main_effects}. This null result aligns with recent evidence that role-based system prompts do not reliably improve LLM task performance~\cite{zheng2024helpful}, and we therefore do not interpret persona as an active driver of the main effects in the remainder of the analysis. As we show in the following subsection, however, even individually neutral or beneficial constraints do not combine additively when imposed together. 

\begin{table}[t]
\caption{Main Effects: Average pass@1 (\%) by Factor Level}
\label{tab:main_effects}
\centering\footnotesize
\begin{tabular}{llccc}
\hline
\textbf{Factor} & \textbf{Model} & \textbf{None} & \textbf{Mid} & \textbf{High} \\
\hline
\multirow{4}{*}{Format} & 4o-mini & 78.1 & 75.4 & 82.0 \\
 & 4o & 83.6 & 81.0 & 86.0 \\
 & 4.1-mini & 87.6 & 88.6 & 88.2 \\
 & 4.1 & 89.2 & 88.2 & 89.1 \\
\hline
\multirow{4}{*}{Persona} & 4o-mini & 79.5 & 78.3 & 77.8 \\
 & 4o & 83.6 & 83.1 & 83.9 \\
 & 4.1-mini & 87.9 & 88.1 & 88.3 \\
 & 4.1 & 89.3 & 88.6 & 88.6 \\
\hline
\multirow{4}{*}{Urgency} & 4o-mini & 76.6 & 78.7 & 80.2 \\
 & 4o & 80.8 & 86.2 & 83.5 \\
 & 4.1-mini & 87.7 & 87.5 & 89.1 \\
 & 4.1 & 88.7 & 88.7 & 89.1 \\
\hline
\multicolumn{5}{l}{\scriptsize Format: None/JSON/XML. Persona: None/Generic/Expert.} \\
\multicolumn{5}{l}{\scriptsize Urgency: None/Moderate/Extreme.} \\
\end{tabular}
\end{table}

\subsection{Super-Additive Degradation in Compound Constraints}

To test whether compound constraints produce degradation beyond the sum of their individual effects, we decompose each triple combination's expected performance using the additive interaction model defined in Eq.~\ref{eq:interaction} of Section~\ref{sec:background}. For each combination of format $f$, persona $p$, and urgency $u$, the predicted pass@1 under additivity is $Y_{\text{base}} + \Delta_f + \Delta_p + \Delta_u$, and the interaction effect $I_{f,p,u}$ is its residual against the actual measured pass@1. Negative values of $I_{f,p,u}$ indicate super-additive degradation, in which the compound constraint produces worse performance than the sum of individual effects would predict.

\begin{figure}[t]
\centering
\includegraphics[width=\columnwidth]{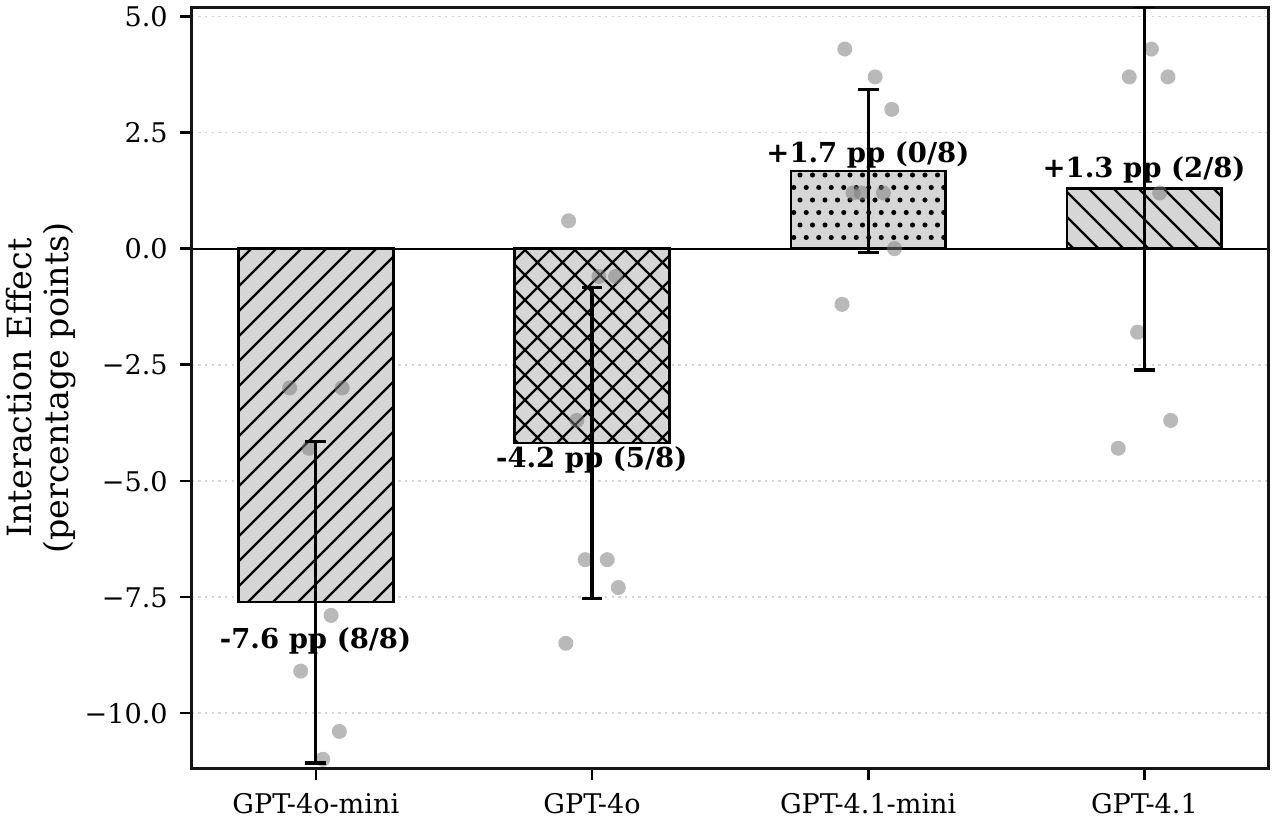}
\caption{Average three-way interaction effect across eight triple-constraint combinations. Negative values indicate super-additive degradation. Numbers in parentheses indicate how many of the eight combinations exceeded a $-$2~pp threshold. The GPT-4o family shows consistent super-additive degradation; the GPT-4.1 family does not.}
\label{fig:interaction}
\end{figure}

Fig.~\ref{fig:interaction} presents the average three-way interaction across all eight triple-constraint combinations for each model. GPT-4o-mini exhibits the strongest super-additive degradation ($\bar{I} = -7.6$~pp, 8/8 combinations super-additive), followed by GPT-4o ($\bar{I} = -4.2$~pp, 5/8). In contrast, GPT-4.1-mini ($\bar{I} = +1.7$~pp, 0/8) and GPT-4.1 ($\bar{I} = +1.3$~pp, 2/8) show no systematic super-additive pattern, with interaction effects near zero or slightly positive. 

\begin{figure}[t]
\centering
\includegraphics[width=\columnwidth]{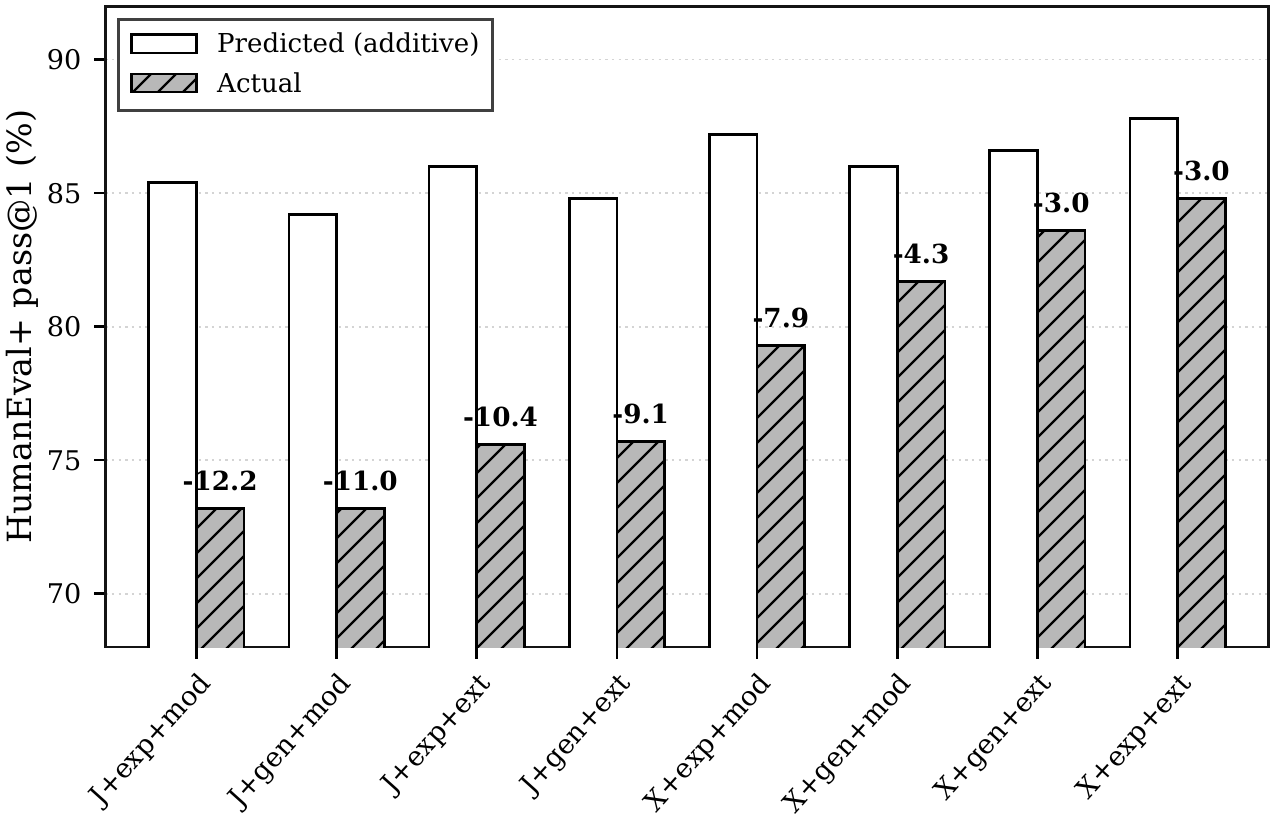}
\caption{Predicted (additive) vs.\ actual pass@1 for all eight triple-constraint combinations on GPT-4o-mini. Every combination underperforms its additive prediction, with JSON combinations showing the largest gaps ($-$9.1 to $-$12.2~pp).}
\label{fig:predicted}
\end{figure}

Fig.~\ref{fig:predicted} provides a detailed breakdown for GPT-4o-mini, the most affected model. All eight triple combinations perform worse than the additive prediction, with interaction effects ranging from $-3.0$~pp (XML+generic+extreme) to $-12.2$~pp (JSON+expert+moderate). JSON-based combinations exhibit substantially larger interaction effects ($\bar{I} = -10.7$~pp) than XML-based combinations ($\bar{I} = -4.6$~pp), suggesting that the computational overhead of JSON string encoding (escape sequences, brace matching) compounds more heavily with concurrent persona and urgency processing than the simpler tag-wrapping structure of XML. Table~\ref{tab:interactions} presents the full interaction decomposition for both GPT-4o family models.

\begin{table}[t]
\caption{Three-Way Interaction Effects (pp) for GPT-4o Family}
\label{tab:interactions}
\centering\footnotesize
\begin{tabular}{lcccccc}
\hline
 & \multicolumn{3}{c}{\textbf{4o-mini}} & \multicolumn{3}{c}{\textbf{4o}} \\
\textbf{Comb.} & Pred & Act & Int & Pred & Act & Int \\
\hline
J+exp+mod & 85.4 & 73.2 & $-$12.2 & 85.4 & 84.8 & $-$0.6 \\
J+gen+mod & 84.1 & 73.2 & $-$11.0 & 85.4 & 84.8 & $-$0.6 \\
J+exp+ext & 86.0 & 75.6 & $-$10.4 & 82.9 & 83.5 & $+$0.6 \\
J+gen+ext & 84.8 & 75.6 & $-$9.1 & 82.9 & 79.3 & $-$3.7 \\
X+exp+mod & 87.2 & 79.3 & $-$7.9 & 94.5 & 87.8 & $-$6.7 \\
X+gen+mod & 86.0 & 81.7 & $-$4.3 & 94.5 & 86.0 & $-$8.5 \\
X+gen+ext & 86.6 & 83.5 & $-$3.0 & 92.1 & 84.8 & $-$7.3 \\
X+exp+ext & 87.8 & 84.8 & $-$3.0 & 92.1 & 85.4 & $-$6.7 \\
\hline
\multicolumn{7}{l}{\scriptsize J=JSON, X=XML, exp=expert, gen=generic, mod=moderate,} \\
\multicolumn{7}{l}{\scriptsize ext=extreme. Pred=Additive prediction, Act=Actual, Int=Interaction.} \\
\end{tabular}
\end{table}

To assess statistical significance of the compound degradation, we conducted McNemar's test on paired per-problem outcomes, comparing conditions that isolate the effect of adding constraints. For GPT-4o-mini, comparing the best single-constraint condition (extreme urgency only, 82.9\%) against the worst triple combination (JSON+expert+moderate, 73.2\%) yielded an odds ratio of 3.67 (95\% CI: [1.49, 9.04], $p = 0.004$), indicating that problems solved under a single constraint were 3.7 times more likely to fail under the triple constraint than vice versa. Similarly, comparing JSON-only (79.3\%) to JSON+expert+moderate (73.2\%) yielded OR~=~2.25 (95\% CI: [0.98, 5.17], $p = 0.076$), a marginally significant trend consistent with the interaction pattern. For GPT-4o, comparing moderate-urgency-only (87.8\%) to JSON+generic+none (75.0\%) was highly significant (OR~=~6.25, 95\% CI: [2.18, 17.96], $p < 0.001$). For GPT-4.1-mini, no pairwise comparison reached significance ($p > 0.05$ for all), consistent with the near-zero interaction effects.

\subsection{Architecture-Dependent Resilience}

The most striking finding is that super-additive degradation is architecture-dependent rather than size-dependent. Within the GPT-4o family, both the smaller model (4o-mini, $\bar{I} = -7.6$ pp) and the larger model (4o, $\bar{I} = -4.2$ pp) exhibit super-additive degradation, although the larger model shows partial resilience. Within the GPT-4.1 family, both models resist the effect entirely ($\bar{I} \approx +1.5$~pp for both). This pattern suggests that the vulnerability to compound constraints is tied to architectural or training differences between model families rather than to parameter count. The GPT-4.1 family was explicitly optimized for instruction-following fidelity~\cite{openai2025gpt41}, and our results indicate that this optimization confers robustness to multi-constraint interference as a secondary benefit.

An important practical implication follows. The effects of prompt constraints cannot be reliably predicted by testing them in isolation. For GPT-4o-mini, each individual constraint either improved performance or had negligible effect, yet their combination produced a 12.2 pp degradation beyond what additivity would predict. This nonlinear interaction represents a latent failure mode that standard single-factor prompt evaluation would not detect.

\subsection{The o3-mini Case}
\label{sec:o3}

The o3-mini reasoning model exhibited a pattern qualitatively distinct from all standard models. Its unconstrained baseline was anomalously low (59.8\%), but performance increased dramatically with any structured output constraint (JSON: 88.6\%, XML: 88.1\%), an improvement of approximately 11~pp. Adding persona and urgency on top of format constraints did not produce further degradation; all 27 conditions with at least one active constraint clustered between 77\% and 92\%. The additive interaction framework is inapplicable to o3-mini because its individual factor effects (20--30~pp each) create ceiling effects that make the three-way interaction term uninterpretable. We attribute this pattern to o3-mini's internal chain-of-thought architecture, which generates hidden reasoning tokens before producing visible output, potentially decoupling reasoning from output formatting. This architectural feature may render structured constraints beneficial rather than taxing, as they provide a concrete output template that complements the model's internal deliberation process.

\subsection{Implications for LLM-Assisted Engineering Workflows}
These results have direct implications for the use of LLMs in software engineering pipelines. In production systems, prompts frequently combine structured output requirements (for downstream parsing), role-based instructions (for domain specialization), and urgency or quality framing (reflecting real-world deployment pressures). Our findings demonstrate that this common practice introduces an interaction-driven failure mode in GPT-4o-class models: compound constraints degrade code generation reliability by 3--12~pp beyond what individual constraint testing would predict.

From a component-level reliability perspective, a 12.2~pp degradation in pass@1 corresponds to approximately 20 additional failing solutions per 164 problems. In automated pipelines where LLM outputs feed directly into test suites or code review, this represents a meaningful increase in defect injection rate. The architecture-dependent nature of this vulnerability suggests that model selection, not just prompt design, is a critical factor in system reliability. Organizations deploying LLM-assisted code generation should evaluate candidate models under compound constraint conditions representative of their actual deployment prompts, rather than relying on single-factor benchmarks.

The asymmetry between JSON and XML interaction effects further suggests a practical mitigation strategy: when structured output is required alongside persona and urgency framing, XML-style wrapping may impose lower compound overhead than JSON string encoding, reducing the magnitude of compound degradation.

To illustrate the broader implications, consider a safety-critical software development workflow in which an LLM assists engineers in implementing functions such as geofence monitoring, battery management, or sensor validation for an unmanned aircraft system. A production prompt may simultaneously require structured JSON output for automated integration, assign an expert developer persona, and include urgency framing during operational troubleshooting. The results presented in this study indicate that evaluating these prompt elements independently may underestimate the probability of introducing logic defects into generated software. Consequently, reliability assessment should be performed using the complete production prompt configuration rather than its individual components.

This framework is intended to complement existing system safety practices rather than replace them. The measured interaction effects can provide additional evidence during software tool assessment, support hazard analyses by identifying prompt configuration as a potential contributor to software defects, and guide verification activities toward prompt configurations representative of operational use. In this way, compound-prompt evaluation extends existing software assurance processes with empirical evidence about an emerging source of reliability degradation in LLM-assisted engineering workflows.

\section{Conclusion}\label{sec:conclusion}
This paper presented a full-factorial empirical study of how structured output formatting, persona assignment, and urgency framing jointly affect the reliability of LLM code generation. Using a $3 \times 3 \times 3$ design across the 164 problems of HumanEval+ and five OpenAI models spanning the GPT-4o family, the GPT-4.1 family, and the o3-mini reasoning model, we produced 22,140 individual evaluations and decomposed each triple-constraint condition into an additive prediction and a residual interaction term, enabling direct measurement of compound effects that cannot be recovered from single-factor experiments.

The results show that the interaction between prompt constraints depends strongly on model architecture rather than parameter count. The GPT-4o family exhibits a consistent super-additive degradation pattern in which compound constraints reduce pass@1 by 3 to 12 percentage points beyond what the sum of individual effects would predict, with the largest interaction reaching $-12.2$ pp on GPT-4o-mini and JSON-based combinations producing substantially larger interaction effects than XML-based combinations. In contrast, the GPT-4.1 family resists the effect at both parameter scales, and the o3-mini reasoning model exhibits a qualitatively different pattern in which structured output constraints improve rather than degrade performance, indicating that explicit instruction-following optimization and internal chain-of-thought generation both appear to confer resilience to multi-constraint interference.

These findings carry two practical implications. The effects of prompt constraints on code generation cannot be reliably extrapolated from single-factor studies, since individually neutral or even beneficial constraints can combine to produce large degradation that standard prompt evaluation would not detect. Model selection, not just prompt design, is therefore a critical factor in deployment reliability, and where structured output is required alongside persona and urgency framing, XML-style wrapping may impose lower compound overhead than JSON string encoding. Important limitations remain: the study covers five OpenAI models on a single benchmark of function-level Python problems under greedy decoding, prompt token length is confounded with constraint intensity, and the additive interaction framework was inapplicable to o3-mini due to ceiling effects.

Future work will extend this framework to additional model families such as Claude, Gemini, and open-weight code models, to additional benchmarks including MBPP+ and repository-level evaluations, and toward mechanistic analysis of why certain architectures resist compound constraints while others do not. A parallel direction is mitigation, exploring whether decoding-time format enforcement, prompt decomposition, or constraint-aware fine-tuning can reduce the magnitude of compound degradation in vulnerable models, moving from documenting compound-prompt failure modes toward predicting and preventing them in deployed LLM-assisted engineering pipelines.


\section*{Data and Code Availability}
The benchmark used in this study is HumanEval+~\cite{liu2024evalplus}, a publicly available EvalPlus project and all model responses were obtained through the OpenAI API using parameters reported in Section~III. Any data and models used in this paper can be made available upon request.

\bibliographystyle{IEEEtran}
\bibliography{references}

\end{document}